# Exploring Sectoral Profitability in the Indian Stock Market Using Deep Learning

**Jaydip Sen**
Department of Data Science and Artificial Intelligence,
Praxis Business School,
Kolkata – 700104, West Bengal, INDIA
Email: jaydip.sen@acm.org

**Bio:** Prof. Jaydip Sen has around 28 years of experience in the field of communication networks protocol design, network analysis, cryptography, network security, and analytics. Currently, he is associated with Praxis Business School, Kolkata, INDIA, as a professor in Data Science and Artificial Intelligence. His research areas include security in wired and wireless networks, intrusion detection systems, secure routing protocols in wireless ad hoc and sensor networks, privacy issues in ubiquitous and pervasive communication and the Internet of Things, machine learning, deep learning, and artificial intelligence. He has published 250 papers in reputed journals and refereed conference proceedings, 22 book chapters, 17 edited volumes and 7 authored books. He also holds five US patents. He has been listed among the top 2% of scientists in the world for the last five consecutive years 2019-2023, as per studies conducted by Stanford University, USA. Prof. Sen serves on the editorial boards of three prestigious international journals and the technical program committees of several international conferences of repute. Prof. Sen is a Senior Member of IEEE, USA, and ACM, USA.

**Hetvi Waghela**
Department of Data Science and Artificial Intelligence,
Praxis Business School,
Kolkata – 700104, West Bengal, INDIA
Email: hetvi.mahendra.waghela_ds23fall@praxistech.school

Hetvi Waghela holds a Bachelor's degree in Computer Science from KJ Somaiya College of Science and Commerce, University of Mumbai, and a Postgraduate Diploma in Data Science from Praxis Tech School. Currently an Associate Data Analyst with 1.2 years of experience in web development, Hetvi has a keen interest in machine learning, deep learning, natural language generation, and artificial intelligence. She has made significant contributions to the research of machine learning in finance, generative AI, and adversarial machine learning. She has published four Scopus-indexed papers and two of her papers are currently under review. She is a member of IEEE, USA, and ACM, USA.

**Sneha Rakshit**
Department of Data Science and Artificial Intelligence,
Praxis Business School,
Kolkata – 700104, West Bengal, INDIA
Email: sneha.rakshit_ds23fall@praxistech.school

Sneha Rakshit earned her BS and MS in Statistics from the University of Kalyani in West Bengal, India, and completed a Postgraduate Diploma in Data Science from Praxis Tech School, Kolkata. She is deeply interested in research areas like machine learning, deep learning, natural language processing, and artificial intelligence. Sneha has published four papers indexed by Scopus, with two more currently under review. Her specific research interests include generative AI and adversarial machine learning. Additionally, she is a member of IEEE and ACM in the USA.

# Exploring Sectoral Profitability in the Indian Stock Market Using Deep Learning

Jaydip Sen[1], Hetvi Waghela[2] and Sneha Rakshit[3]
Praxis Business School, Kolkata, India
email: [1]jaydip.sen@acm.org, {[2]hetvi.mahendra.waghela_ds23fall, [3]sneha.rakshit_ds23fall}@praxistech.school

**Abstract:** This paper explores using a deep learning Long Short-Term Memory (LSTM) model for accurate stock price prediction and its implications for portfolio design. Despite the efficient market hypothesis suggesting that predicting stock prices is impossible, recent research has shown the potential of advanced algorithms and predictive models. The study builds upon existing literature on stock price prediction methods, emphasizing the shift toward machine learning and deep learning approaches. Using historical stock prices of 180 stocks across 18 sectors listed on the NSE, India, the LSTM model predicts future prices. These predictions guide buy/sell decisions for each stock and analyze sector profitability. The study's main contributions are threefold: introducing an optimized LSTM model for robust portfolio design, utilizing LSTM predictions for buy/sell transactions, and insights into sector profitability and volatility. Results demonstrate the efficacy of the LSTM model in accurately predicting stock prices and informing investment decisions. By comparing sector profitability and prediction accuracy, the work provides valuable insights into the dynamics of the current financial markets in India.

## 1. Introduction

Designing predictive models for accurately forecasting future stock prices has always been a fascinating and challenging endeavor, given the volatile and unpredictable nature of stock markets. Despite the efficient market hypothesis suggesting that such predictions are impossible, there are propositions in the literature demonstrating that advanced algorithms and predictive models can effectively forecast future stock prices. Two of the oldest methods for stock price prediction are time series decomposition and exponential smoothing (Sen, 2018c). Recently, the use of machine learning and deep learning systems has become the most popular approach (Sen, 2018a; Bollen et al., 2011; Mehtab & Sen, 2020a; Sen et al., 2020). Bollen et al. illustrate how emotions expressed on the social web can significantly impact stock market transaction volumes (Bollen et al., 2011). Additionally, some studies have highlighted the effectiveness of *convolutional neural networks* (CNN) in accurately forecasting stock prices (Mehtab & Sen, 2020a).

The technical analysis of stocks, aimed at forecasting future prices is widely recognized. Researchers have proposed various approaches to technical analysis, which involve identifying established patterns in the stock price time series and devising investment strategies to generate profits (Sen et al., 2024). A wide range of well-defined patterns and indicators have been suggested in the literature for this purpose.

The present study introduces a deep learning-based *long-and-short-term memory* (LSTM) model designed to accurately predict future stock prices. This model automatically retrieves historical stock prices using a Python function, utilizing stock ticker names from the NSE within a specified interval determined by a start and end date. Utilizing the historical prices of 180 stocks across 18 sectors, the model predicts future stock prices. Using these predictions, buy/sell decisions are made for each stock, and ultimately, the total profit for each stock is calculated. By aggregating the profits of all stocks within a sector, the overall profitability of the sector is determined. A comparative analysis is then conducted to evaluate the profitability of the eighteen sectors and the prediction accuracy of the LSTM model.

The current work makes three main contributions. Firstly, it introduces an optimized deep-learning model that leverages the LSTM architecture to predict future stock prices, aiding in robust portfolio design. Secondly, the LSTM predictions are utilized to inform buy/sell transactions involving 180 stocks selected from eighteen sectors listed on the NSE, India. The high precision of the model demonstrates its efficacy and effectiveness. Thirdly, the stock returns provide insights into the current profitability of investments and the volatilities within the eighteen sectors.

The paper is organized as follows. Section 2 briefly discusses recent works in the literature on stock price prediction and portfolio design. Section 3 presents data extraction, preprocessing, and the methodology used in the current work. Section 4 discusses the design of the proposed LSTM model for stock price prediction. Detailed results and their analysis are presented in Section 5. Finally, Section 6 concludes the paper.

## 2. Related Work

Researchers have extensively tackled the challenges of accurately predicting future stock prices and creating optimal portfolios that balance risks and returns. Many methods have been proposed, including time series decomposition, statistical and econometric modeling, such as ARIMA, ARDL, and VAR, to forecast future stock prices. (Chatterjee, et al., 2021; Du, 2018; Khandelwal et al., 2015; Ning et al., 2019; Pai & Lin, 2015; Sen, 2022d; Sen, 2018b; Sen, 2018c; Sen, 2017a; Sen, 2017b; Sen & Datta Chaudhuri, 2017a; Sen & Datta Chaudhuri, 2017b; Sen & Datta Chaudhuri, 2017c; Sen & Datta Chaudhuri, 2018; Sen & Datta Chaudhuri, 2016a; Sen & Datta Chaudhuri, 2016b; Sen & Datta Chaudhuri, 2016c; Sen & Datta Chaudhuri, 2016d; Sen & Datta Chaudhuri, 2015; Wang et al., 2020).

Learning-based algorithms and architectures have been widely employed to improve the accuracy of stock price predictions and to strengthen the resilience of stock portfolios (Ballings et al., 2015; Bao et al., 2017; Binkowski et al., 2018; Lv et al., 2019; Mehtab & Sen, 2022; Mehtab & Sen, 2021; Mehtab & Sen, 2020a; Mehtab & Sen, 2020b; Mehtab et al., 2021; Mehtab et al., 2020; Sen, 2018a; Sen & Datta Chaudhuri, 2017a; Sen & Mehtab, 2022b; Sen & Mehtab, 2021a; Sen & Mehtab, 2021b; Sen et al., 2021c; Sen et al., 2021g; Sen et al., 2020; Wu et al., 2012; Yang et al., 2017).

Numerous hybrid models proposed in the literature combine learning algorithms with sentiment information from social media to generate more accurate predictions (Attigeri et al., 2015; Chen et al., 2019; Galvez & Gravano, 2017; Medhat et al. 2014; Mehtab & Sen, 2019; Mittal & Goel, 2012; Nam & Seong, 2019; Jing et al., 2019; Shi et al., 2019; Weng et al., 2017).

Various methods, including metaheuristics, risk parity, principal component analysis, different versions of mean-variance optimization, reinforcement learning, and pair-trading have been suggested for creating optimal portfolios (Almahdi et al., 2019; Chen & Zhou, 2018; Ertenlice & Kalayci, 2018; Macedo et al, 2017; Pai, 2017; Petchrompo et al., 2021; Qu et al., 2017; Reveiz-Herault, 2016; Sen, 2023; Sen, 2022a; Sen, 2022b; Sen, 2022c; Sen, 2023; Sen, 2022a; Sen & Dutta, 2022a; Sen & Dutta, 2021; Sen & Dutta, 2022b; Sen et al., 2021c; Sen & Mehtab, 2022a; Sen et al., 2021d; Sen et al., 2021e; Sen et al., 2021f; Sen & Sen, 2023; Sen et al., 2024; Wang et al., 2022).

The use of generalized autoregressive conditional heteroscedasticity (GARCH) for estimating and predicting future volatility of stock prices in the Indian stock market is prosed in some work (Sen et al., 2021b).

The current study proposes the utilization of a deep-learning regression model based on the LSTM architecture to accurately predict future stock prices. Using the model's predictions, decisions to buy or sell are made, and the resulting profits from these transactions are aggregated for multiple stocks chosen from eighteen critical sectors listed on the NSE, India. The profitability of each stock is measured by the gross ratio of profit earned over a specified period to the mean price of the stock over the same period. This ratio is used as a metric to gauge profitability. The average of this ratio for stocks within the same sector serves as a measure of the overall sector's profitability. Various sectors are analyzed, and their profitability measures are calculated. The results provide valuable insights into sector profitability for potential investors in the Indian stock market. Furthermore, the study demonstrates the effectiveness and efficiency of the predictive model.

## 3. Data and Methodology

The steps are as follows. The approach taken in this study consists of five primary stages. These stages are as follows: (1) Data acquisition and loading, (2) Model design, (3) Processing the model output, (4) Visualizing the output through plotting, and (5) Predicting future stock prices. The following sections will provide a brief overview of each method.

**(1) Data acquisition and loading:** This stage involves extracting historical stock data from the Yahoo Finance website and performing the required preprocessing and transformation of the raw data. Stock prices are retrieved using their ticker names, start date, and end date. Python libraries are utilized to create functions for this task. The start date for all stocks is set as January 1, 2005, and the end date as April 23, 2024.

**(2) Model Design:** In this phase, the LSTM-based predictive model is formulated. A function is employed to design the model, utilizing various parameters: (i) the length of historical stock price data used as input, (ii) the number of variables (features) utilized in the input, (iii) the number of nodes in the LSTM layer. (iv) the number of layers in the model, (v) the percentage of nodes used in dropout for regularization, (vi) the type of loss function used for training and validation, (vii) the optimizer employed, (viii) the batch size for training the model, and (ix) the number of epochs for model training.

All these parameters are adjustable. However, the following values have been applied during the training and validation phases:

The input data's historical length is set to 50, indicating the utilization of the past 50 days' stock price records for making predictions. Five input features are utilized: open, high, low, close, and volume. The "close" variable is the prediction target, while the others are explanatory variables, also known as predictor variables. The default number of LSTM nodes in the model is 256, and the default number of LSTM layers is two. To ensure proper regularization of the model, the dropout percentage is set to 30% by default. The Huber loss function is chosen due to its adaptiveness in handling complex data. The Adam optimizer is employed as the default optimizer for optimizing the gradient descent algorithm during learning. A batch size of 64 and 100 epochs are used as defaults for training the LSTM model.

**(3) Processing the output of the model:** In this stage, a Python function is executed with two arguments: (i) the model designed in the previous step, and (ii) the transformed and pre-processed data. This function undertakes several core tasks based on the actual and predicted stock prices.

When the model predicts a higher price for the next day compared to today's price, it advises the investor to buy the stock. However, buy profits are calculated based on the actual price of the next day, not the model's previous-day prediction. The total buy profit for the stock is determined by summing the buy profits for all days when the stock was sold.

Conversely, if the model predicts a lower price for the next day, a sell strategy is recommended for the current day. The selling profit is computed as the difference between today's actual stock price and the actual price of the stock the next day. The total sell profit for the stock is calculated by summing the sell profits for all days when the stock was bought.

The total profit is obtained by adding the total buy profit to the total sell profit.

To assess the model's prediction accuracy, this function also calculates two metrics: mean absolute error and accuracy score. The mean absolute error represents the average of the absolute differences between the actual and predicted prices, while the accuracy score indicates the percentage of cases where the LSTM model correctly predicts the direction (up or down) of the next day's price compared to the current day.

In this study, it is assumed that a fictional investor engages in mandatory trading every day. If the forecasted price for the next day is higher than the current day's price, the investor must buy the stock. Conversely, if the forecasted price is lower or equal to the current day's price, the investor must sell the stock. Buy/sell profits are computed assuming the trading of a single equity share. However, the returns for stocks and sectors are independent of the number of equity shares traded, as they are presented as percentage figures.

**(4) Visualizing the output:** This step involves creating various visualizations. Key plots include training and validation loss plots across different epochs, as well as plots comparing actual and predicted stock prices generated by the LSTM.

**(5) Forecasting future stock prices:** This stage entails utilizing a function to calculate the future values of stock prices by executing the LSTM model. The forecast horizon, which determines how far into the future predictions are made, can be adjusted in the function. However, in this study, a forecast horizon of one day is employed, predicting the stock price for the following day.

## 4. The Deep Learning Model Architecture

In the previous section, we discussed how stock prices are predicted using a one-day forecast horizon with the assistance of a Long Short-Term Memory (LSTM) model. This section aims to provide a detailed insight into the architecture of the model and the rationale behind selecting various parameters.

LSTM, a type of recurrent neural network (RNN), is particularly adept at handling sequential data such as time series of stock prices or text (Geron, 2019). Unlike traditional RNNs, LSTM networks maintain long-term dependencies in the data through special mechanisms called gates, allowing them to effectively capture patterns and predict future values.

For our stock price prediction task, we have meticulously designed and fine-tuned an LSTM model. Let's break down the architecture:

*Input Data*: We use the daily close prices of the stock for the past 50 days as our input. This input data, consisting of a single feature (close values), is shaped as (50, 1).

*First LSTM Layer*: The input data is fed into the first LSTM layer, which comprises 256 nodes. Each record in the input generates 256 features from the LSTM nodes, resulting in an output shape of (50, 256).

*Dropout Layer*: To prevent overfitting, we include a dropout layer after the first LSTM. This layer randomly deactivates thirty percent of the nodes' output.

*Second LSTM Layer*: Another LSTM layer, identical to the previous one, receives the output from the first LSTM layer and applies a dropout rate of thirty percent.

*Dense Layer*: A dense layer with 256 nodes processes the output from the second LSTM, yielding the predicted close price.

*Forecast Horizon*: We can adjust the forecast horizon by modifying a tunable parameter. In our case, a forecast horizon of one day is used to predict the price for the following day.

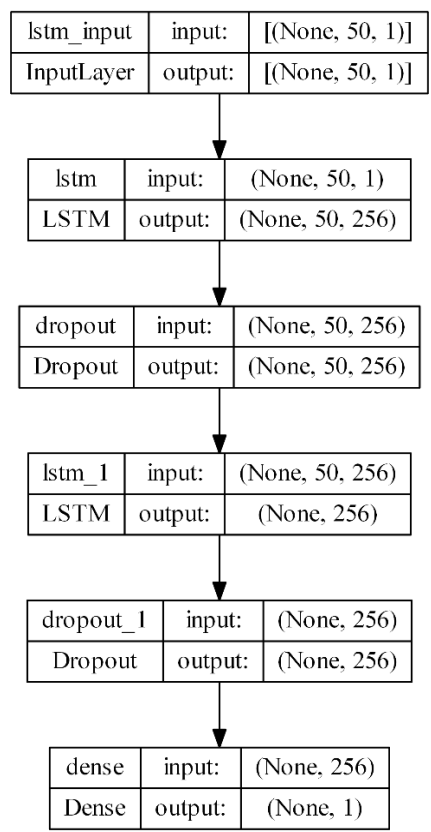

**Figure 1.** The LSTM model for daily stock price prediction.

*Training Parameters*: The model is trained with a batch size of 64 and 100 epochs. We employ the ReLU activation function for all layers except the final output layer, where the sigmoid function is used. This choice of activation functions helps in handling non-linearities in the data.

*Loss and Accuracy Measurement*: To evaluate the model's performance, we utilize the Huber loss function to measure the loss and the mean absolute error function to measure accuracy. The Huber loss function is preferred due to its ability to combine the characteristics of Mean Squared Error (MSE) and Mean Absolute Error (MAE), providing a robust evaluation metric (Geron, 2019).

Hyperparameter Optimization: The model's hyperparameters are optimized using the grid search method, ensuring the selection of optimal values.

By adopting this comprehensive approach, we aim to build a robust LSTM model capable of accurately predicting stock prices, leveraging its ability to capture intricate patterns in sequential data.

## 5. Experimental Results

Eighteen significant sectors are selected from the list of stock sectors on the NSE, India. These sectors are (i) *auto*, (ii) *banking*, (iii) *consumer durables*, (iv) consumption (v) energy, (vi) ESG, (vii) financial services, (viii) fast-moving *consumer goods* (FMCG), (ix) infrastructure, (x) *information technology*, (xi) *media*, (xii) *metal*, (xiii) *oil & gas*, (xiv) pharma (xv) private banks, (xvi) PSU banks, (xvii) *realty*, and (xviii) services. Based on the NSE's report from March 28, 2024, the ten most significant stocks in these sectors are identified (NSE Website). These stocks have the largest free-float market capitalization values in their respective sectors, and hence they contribute most significantly to their respective sectoral index calculations. The LSTM model is employed to calculate the total profit (sum of buy and sell profits) for each stock. The average ratio of total profit to the mean price of each stock during the study period (January 1, 2005 – April 24, 2024) serves as the sector's profitability metric. The models are developed using Python libraries and executed on a laptop computer with an Intel i7-9750H CPU, 4.46 GHz based clock speed, 32 GB RAM, and 6GB memory of the GPU. The operating system of the machine is 64-bit Windows 11. Training the models for one epoch took an average of 15 seconds. The models undergo 100 epochs of training. The results for the eighteen sectors are detailed below.

*Auto sector:* Based on the report published by the NSE on March 24, 2024, the ten most significant stocks in the *auto* sector listed in NSE and their weights (in percentage) in computing the sector index are as follows. Mahindra & Mahindra (M&M): 18.26, Tata Motors (TATAMOTORS): 17.35, Maruti Suzuki India (MARUTI): 16.51, Bajaj Auto (BAJAJ-AUTO): 10.28, Hero MotoCorp (HEROMOTOCO): 6.09, Eicher Motors (EICHERMOT): 5.46, TVS Motor Company (TVSMOTOR): 4.97, Tata Motors DVR (TATAMTRDVR): 3.05, Bharat Forge (BHARATFORG): 2.87, and Samvardhana Motherson International (MOTHERSON): 2.76 (NSE Website). Table 1 displays the outcomes for auto sector stocks, detailing the total profits from buying and selling each stock between January 1, 2005, and April 23, 2024. These profits are normalized against each stock's mean price to provide a comparable measure. The sector's overall profitability is determined by averaging the ratios of total profit to mean price across all ten stocks in the sector. Figure 2 illustrates the training and validation loss of the LSTM model over 100 epochs for the sector's leading stock, M&M, based on its free-float market capitalization. Additionally, Figure 3 showcases the actual and predicted prices of M&M stock by the LSTM model from January 1, 2024, to April 23, 2024.

**Table 1.** The results of the Auto sector stocks (Period: January 1, 2005 – April 23, 2024)

| Stock | Buy Profit | Sell Profit | Gross Profit | Mean Price | Gross Profit / Mean Price |
|---|---|---|---|---|---|
| M&M | 288537 | 276382 | 564919 | 369 | 1530.95 |
| TATAMOTORS | 163784 | 162912 | 326696 | 180 | 1814.98 |
| MARUTI | 1768834 | 1775983 | 3544817 | 3578 | 990.73 |
| BAJAJ-AUTO | 819358 | 833160 | 1652518 | 1647 | 1003.35 |
| HEROMOTOCO | 628732 | 628360 | 1257092 | 1585 | 793.12 |
| EICHERMOT | 877605 | 860291 | 1737896 | 1410 | 907.49 |
| TVSMOTOR | 201876 | 202474 | 404349 | 262 | 1543.31 |
| TATAMTRDVR | 45980 | 45555 | 91535 | 191 | 479.24 |
| BHARATFORG | 210759 | 209443 | 320202 | 263 | 1597.72 |
| MOTHERSON | 19950 | 19457 | 39407 | 32 | 1231.47 |
| **Sectoral average of gross profit/mean price** | | | | | **1189.24** |

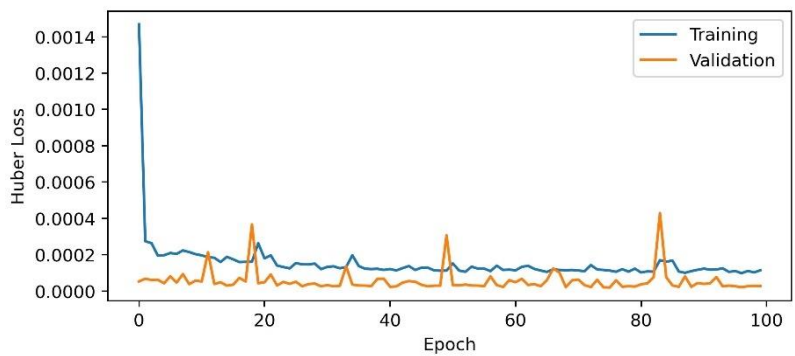

**Figure 2.** The loss of the LSTM model for the M&M stock

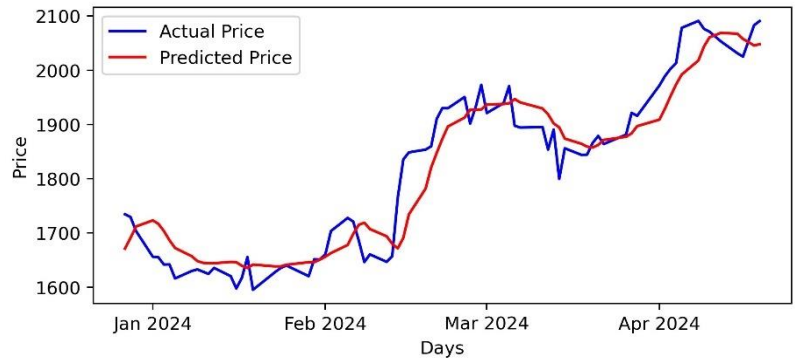

**Figure 3.** The actual vs predicted price of the M&M stock from January 1, 2024, to April 23, 2024

*Banking sector:* As per the report published by the NSE on March 28, 2024, the ten stocks with the highest free-float market capitalization in the *banking* sector and their respective weights (in percentage) in computing the overall sectoral index are as follows. HDFC Bank (HDFCBANK): 29.00, ICICI Bank (ICICIBANK): 23.73, Axis Bank (AXISBANK): 9.19, State Bank of India (SBIN): 9.14, Kotak Mahindra Bank (KOTAKBANK): 9.11, IndusInd Bank (INDUSINDBK): 6.92, Bank of Baroda (BANKBARODA): 3.31, Punjab National Bank (PNB): 2.49, Federal Bank (FEDERALBNK): 2.27, and IDFC First Bank (IDFCFIRSTB): 2.01 (NSE Website). The results of this sector are presented in Table 2.

**Table 2.** The results of the Banking sector stocks (Period: January 1, 2005 – April 23, 2024)

| Stock | Buy Profit | Sell Profit | Gross Profit | Mean Price | Gross Profit / Mean Price |
|---|---|---|---|---|---|
| HDFCBANK | 364131 | 370223 | 734354 | 433 | 1695.97 |
| ICICIBANK | 140462 | 132266 | 272728 | 271 | 1006.38 |
| AXISBANK | 208270 | 209023 | 417293 | 315 | 1324.74 |
| SBIN | 120565 | 118709 | 239274 | 175 | 1367.28 |
| KOTAKBANK | 401103 | 399902 | 801005 | 651 | 1230.42 |
| INDUSINDBK | 337654 | 330735 | 668389 | 571 | 1170.56 |
| BANKBARDA | 34257 | 32601 | 66858 | 95 | 703.77 |
| PNB | 30871 | 32024 | 62895 | 91 | 691.15 |
| FEDERALBNK | 29606 | 29812 | 59418 | 39 | 1523.54 |
| IDFCFIRSB | 3127 | 3374 | 6501 | 51 | 127.47 |
| **Sectoral average of gross profit/mean price** | | | | | **1084.13** |

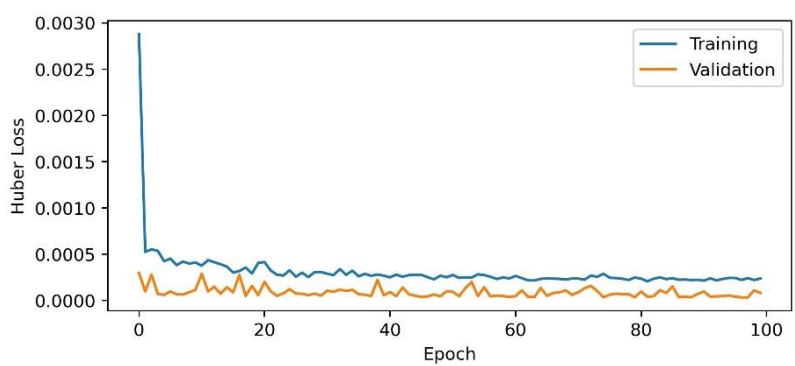

**Figure 4.** The loss of the LSTM model for the HDFCBANK stock

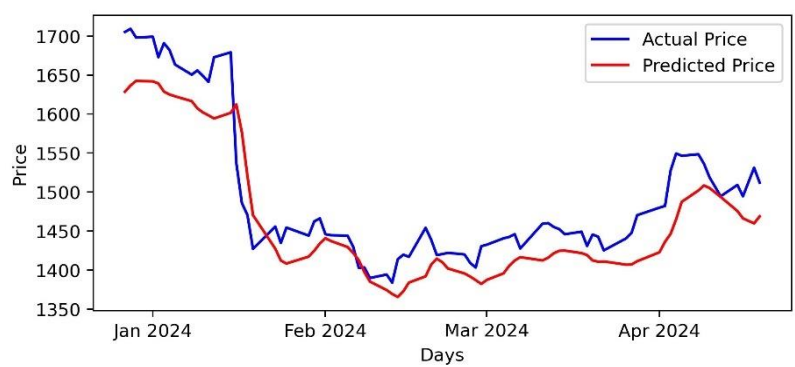

**Figure 5.** The actual vs predicted price of the HDFCBANK stock from January 1, 2024, to April 23, 2024

The training and validation loss for the LSTM model for HDFC Bank, the leading stock of this sector, are plotted in Figure 4. The actual and predicted prices of the HDFC Bank stock by the LSTM model for the period January 1, 2024, and April 23, 2024, are depicted in Figure 5.

*Consumer Durables sector:* Based on the NSE's report published on April 23, 2024, the top ten stocks of this sector based on their free-float market capitalization values and their respective weights (in percentage) in the computation of the overall sectoral index are as follows. Titan Company (TITAN): 33.06, Havells India (HAVELLS): 14.06, Dixon Technologies India (DIXON): 10.44, Voltas (VOLTAS): 9.33, Crompton Greaves Consumer Electricals (CROMPTON): 6.37, Blue Star (BLUESTARCO): 6.11, Kalyan Jewellers India (KALYANKJIL): 3.59, Kajaria Ceramics (KAJARIACER): 3.54, Bata India (BATAINDIA): 3.25, and Amber Enterprises India (AMBER): 2.46 (NSE Website). The results of this sector are presented in Table 3.

**Table 3.** The results of the Consumer Durables sector stocks **(**Period: January 1, 2005 – April 24, 2024)

| Stock | Buy Profit | Sell Profit | Gross Profit | Mean Price | Gross Profit / Mean Price |
|---|---|---|---|---|---|
| TITAN | 498158 | 510217 | 1008375 | 505 | 1996.78 |
| HAVELLS | 227281 | 226364 | 453645 | 339 | 1338.19 |
| DIXON | 395398 | 369159 | 764557 | 2867 | 266.67 |
| VOLTAS | 198374 | 197248 | 395622 | 331 | 1195.23 |
| CROMPTON | 16576 | 15977 | 32553 | 276 | 117.95 |
| BLUESTARCO | 121653 | 115002 | 236655 | 217 | 1090.58 |
| KALYANKJIL | 7960 | 9016 | 16976 | 163 | 51.29 |
| KAJARIACER | 218245 | 215072 | 433317 | 330 | 1313.08 |
| BATAINDIA | 349392 | 336876 | 686268 | 573 | 1197.68 |
| AMBER | 180353 | 170320 | 350673 | 2108 | 166.35 |
| **Sectoral average of gross profit/mean price** | | | | | **873.38** |

The training and validation loss for the LSTM model for Titan Company, the leading stock of the consumer durables sector, are plotted in Figure 6. The actual and predicted prices of the Titan Company stock by the LSTM model for the period January 1, 2024, and April 23, 2024, are depicted in Figure 7.

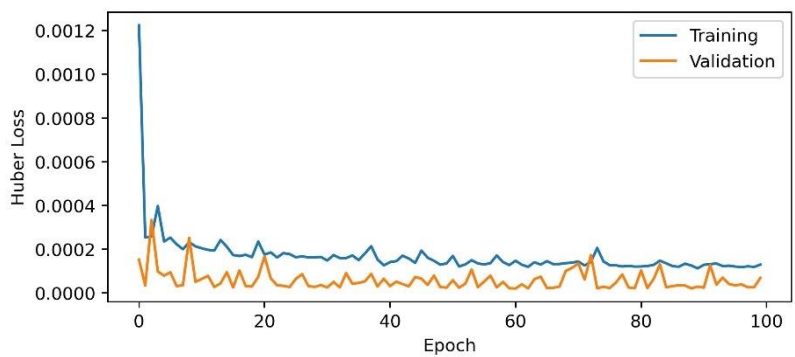

**Figure 6.** The loss of the LSTM model for the TITAN stock

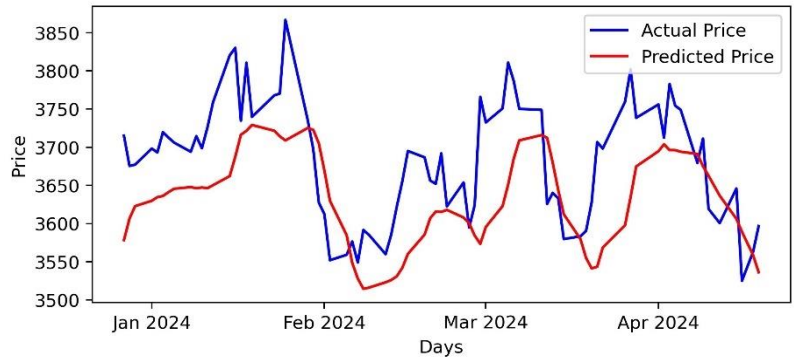

**Figure 7.** The actual vs predicted price of the TITAN stock from January 1, 2024, to April 23, 2024

*FMCG sector:* Based on the NSE's report published on April 23, 2024, the top ten stocks of this sector based on their free-float market capitalization values and their respective weights (in percentage) in the computation of the overall sectoral index are as follows. ITC (ITC): 32.88, Hindustan Unilever (HINDUNILVR): 19.14, Nestle India (NESTLEIND): 8.86, Tata Consumer Products (TATACONSUM): 6.53, Varun Beverages (VBL): 6.36, Britannia Industries (BRITANNIA): 5.49, Godrej Consumer Products (GODREJCP): 4.48, Colgate Palmolive India (COLPAL): 3.42, United Spirits (MCDOWELL-N): 3.12, and Dabur India (DABUR): 2.90 (NSE Website). The results of this sector are presented in Table 4.

**Table 4.** The results of the FMCG sector stocks (Period: January 1, 2005 – April 23, 2024)

| Stock | Sell Profit | Buy Profit | Gross Profit | Mean Price | Gross Profit / Mean Price |
|---|---|---|---|---|---|
| ITC | 82670 | 84128 | 166798 | 114 | 1463.15 |
| HINDUNILVR | 548689 | 561014 | 1109703 | 698 | 1589.83 |
| NESTLEIND | 2806609 | 2951173 | 5757782 | 5518 | 1043.45 |
| TATACONSUM | 165085 | 160608 | 325693 | 181 | 1799.41 |
| VBL | 67004 | 65846 | 132850 | 343 | 387.32 |
| BRITANNIA | 980788 | 984330 | 1965118 | 1040 | 1889.54 |
| GODREJCP | 199040 | 200956 | 399996 | 357 | 1120.44 |
| COLPAL | 326193 | 344037 | 670230 | 674 | 994.41 |
| MCDOWELL-N | 172566 | 166815 | 339381 | 397 | 854.86 |
| DABUR | 113178 | 112972 | 226150 | 211 | 1071.80 |
| **Sectoral average of gross profit/mean price** | | | | | **1221.42** |

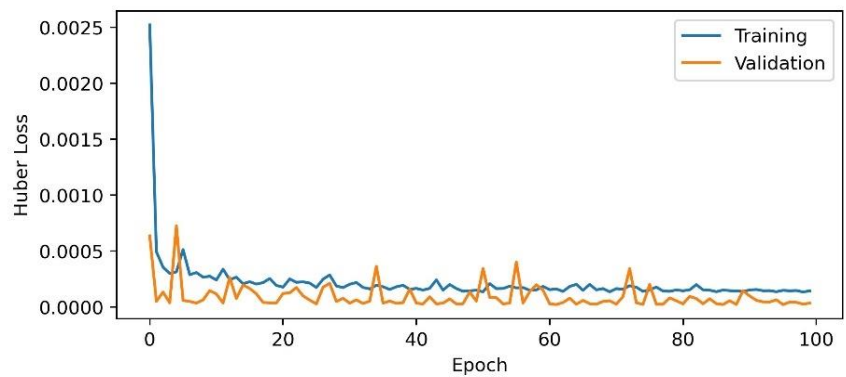

**Figure 8.** The loss of the LSTM model for the ITC stock

The training and validation loss for the LSTM model for ITC, the leading stock of this sector, are plotted in Figure 8. The actual and predicted prices of the ITC stock by the LSTM model from January 1, 2024, to April 23, 2024, are depicted in Figure 9.

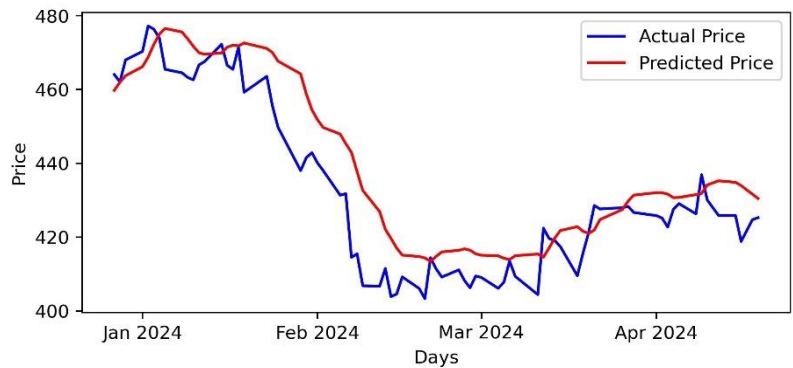

**Figure 9.** The actual vs predicted price of the ITC stock from January 1, 2024, to April 23, 2024

**Table 5.** The results of the Pharma sector stocks (Period: January 1, 2005 – April 23, 2024)

| Stock | Buy Profit | Sell Profit | Gross Profit | Mean Price | Gross Profit / Mean Price |
|---|---|---|---|---|---|
| SUNPHARMA | 253891 | 261400 | 515291 | 328 | 1571.01 |
| CIPLA | 246238 | 239764 | 486002 | 358 | 1357.55 |
| DRREDDY | 1242154 | 1230867 | 2473021 | 1729 | 1430.32 |
| DIVISLAB | 722230 | 680952 | 1403182 | 1156 | 1214.88 |
| LUPIN | 309512 | 314236 | 623748 | 601 | 1037.85 |
| AURPHARMA | 236404 | 243760 | 480164 | 273 | 1758.84 |
| ZYDUSLIFE | 114524 | 112443 | 226967 | 199 | 1140.54 |
| ALKEM | 214841 | 208592 | 423433 | 2558 | 165.53 |
| TORNTPHARMA | 315600 | 311716 | 627316 | 488 | 1285.48 |
| IPCALAB | 236607 | 232129 | 468736 | 263 | 1782.27 |
| **Sectoral average of gross profit/mean price** | | | | | **1274.43** |

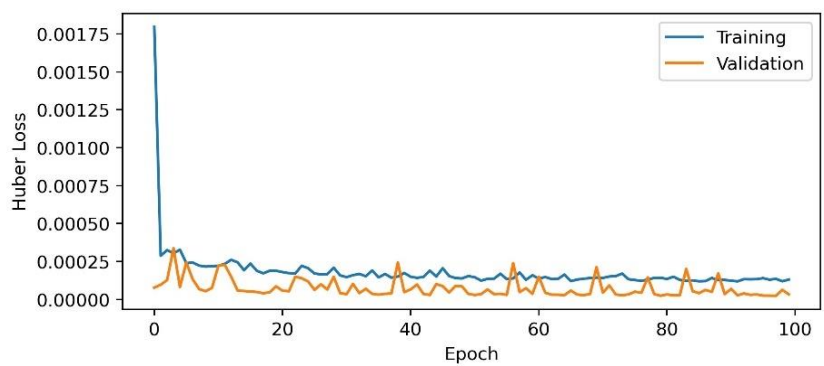

**Figure 10.** The loss of the LSTM model for the SUNPHARMA stock

*Pharma sector:* Based on the NSE's report published on March 28, 2024, the top ten stocks of this sector based on their free-float market capitalization values and their respective weights (in percent) in the computation of the overall sectoral index are as follows. Sun Pharmaceuticals Industries (SUNPHARMA): 26.85, Cipla (CIPLA): 12.06, Dr. Reddy's Laboratories

(DRREDDY): 11.51, Divi's Laboratories (DIVISLAB): 6.74, Lupin (LUPIN): 5.99, Aurobindo Pharma (AURPHARMA): 4.70, Zydus Lifesciences (ZYDUSLIFE): 3.91, Alkem Laboratories (ALKEM): 3.81, Torrent Pharmaceuticals (TORNTPHARM): 3.65, and Ipca Laboratories (IPCALAB): 2.55 (NSE Website). The results of this sector are presented in Table 5.

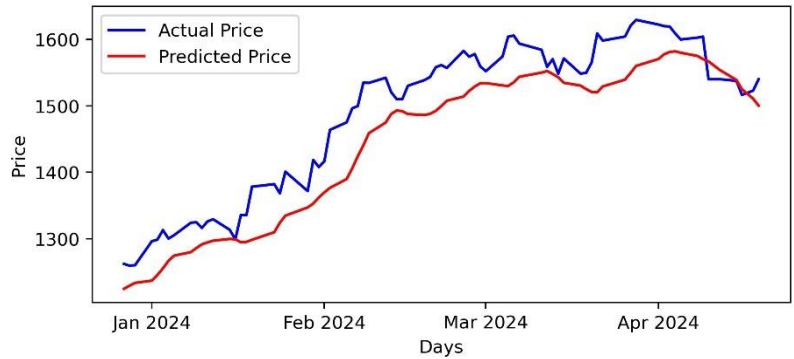

**Figure 11.** The actual vs predicted price of the SUNPHARMA stock from January 1, 2024, to April 23, 2024

The training and validation loss for the LSTM model for Sun Pharmaceuticals, the leading stock of this sector, are plotted in Figure 10. The actual and predicted prices of the Sun Pharmaceuticals stock by the LSTM model from January 1, 2024, to April 23, 2024, are depicted in Figure 11.

*Information Technology (IT) sector:* Based on the NSE's report published on March 28, 2024, the top ten stocks of this sector based on their free-float market capitalization values and their respective weights (in percentage) in the computation of the overall sectoral index are as follows. Infosys (INFY): 26.80, Tata Consultancy Services (TCS): 24.86, HCL Technologies (HCLTECH): 10.34, Tech Mahindra (TECHM): 9.97, Wipro (WIPRO): 8.52, LTIMindtree (LTIM): 5.70, Persistent Systems (PERSISTENT): 5.17, Coforge (COFORGE): 4.24, MphasiS (MPHASIS): 2.50, and L&T Technology Services (LTTS): 1.90 (NSE Website). The results of this sector are presented in Table 6.

**Table 6.** The results of the IT sector stocks (Period: January 1, 2005 – April 23, 2024)

| Stock | Buy Profit | Sell Profit | Gross Profit | Mean Price | Gross Profit / Mean Price |
|---|---|---|---|---|---|
| INFY | 295161 | 308620 | 603781 | 382 | 1580.58 |
| TCS | 610753 | 611463 | 1222216 | 1093 | 1118.22 |
| HCLTECH | 205669 | 194262 | 399931 | 318 | 1257.64 |
| TECHM | 170839 | 164743 | 335582 | 473 | 709.48 |
| WIPRO | 96302 | 102070 | 198372 | 158 | 1255.52 |
| LTIM | 377392 | 379652 | 757044 | 2930 | 258.38 |
| PERSISTENT | 287332 | 280026 | 567358 | 686 | 827.05 |
| COFORGE | 720249 | 699972 | 1420221 | 1103 | 1287.60 |
| MPHASIS | 403666 | 379648 | 783314 | 648 | 1208.82 |
| LTTS | 324679 | 296236 | 620915 | 2497 | 248.66 |
| **Sectoral average of gross profit/mean price** | | | | | **975.20** |

The training and validation loss for the LSTM model for Infosys, the leading stock of this sector, are plotted in Figure 12. The actual and predicted prices of the Infosys stock by the LSTM model from January 1, 2024, to April 23, 2024, are depicted in Figure 13.

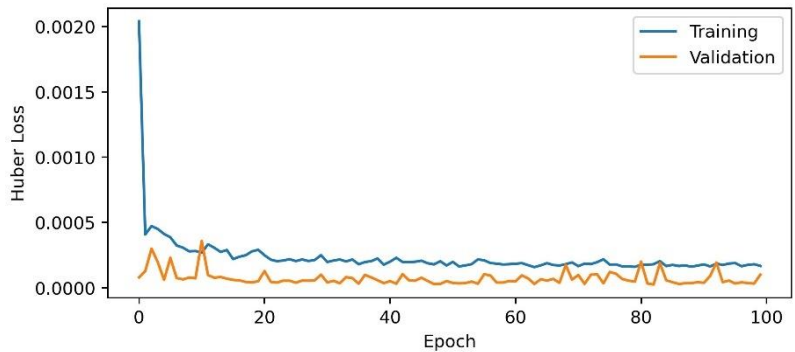

**Figure 12.** The loss of the LSTM model for the INFY stock

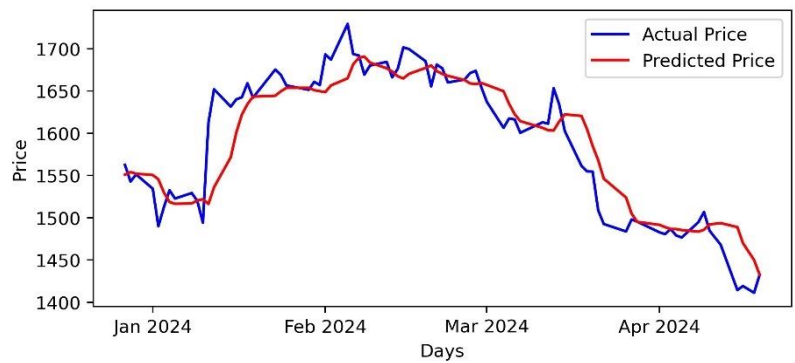

**Figure 13.** The actual vs predicted price of the INFY stock from January 1, 2024, to April 23, 2024

**Table 7.** The results of the Media sector stocks (Period: January 1, 2005 – April 23, 2024)

| Stock | Buy Profit | Sell Profit | Gross Profit | Mean Price | Gross Profit / Mean Price |
|---|---|---|---|---|---|
| ZEEL | 88564 | 87606 | 176170 | 199 | 885.28 |
| PVRINOX | 5034 | 4723 | 9757 | 1557 | 6.27 |
| SUNTV | 65723 | 71781 | 137504 | 350 | 392.87 |
| TV18BRDCST | 12359 | 13252 | 25611 | 50 | 512.22 |
| SAREGAMA | 48530 | 49169 | 97699 | 64 | 1526.55 |
| NAZARA | 14965 | 14981 | 29946 | 763 | 39.25 |
| DISHTV | 13327 | 13021 | 26348 | 49 | 537.71 |
| NETWORK18 | 67959 | 70309 | 138268 | 127 | 1088.72 |
| TIPSINDLTD | 25531 | 26727 | 52258 | 30 | 1741.93 |
| HATHWAY | 5266 | 5391 | 10657 | 34 | 313.44 |
| **Sectoral average of gross profit/mean price** | | | | | **704.42** |

*Media sector:* Based on the NSE's report published on March 28, 2024, the top ten stocks of this sector based on their free-float market capitalization values and their respective weights (in percentage) in the computation of the overall sectoral index are as follows. Zee Entertainment Enterprises (ZEEL): 29.67, PVR INOX (PVRINOX): 21.76, Sun TV Network (SUNTV): 11.54, TV18 Broadcast (TV18BRDCST): 7.78, Saregama India (SAREGAMA):

6.38, Nazara Technologies (NAZARA): 5.54, Dish TV India (DISHTV): 5.43, Network18 Media & Investments (NETWORK18): 5.29, Tips Industries (TIPSINDLTD): 4.48, and Hathway Cable & Datacom (HATHWAY): 2.12 (NSE Website). The results of this sector are presented in Table 7.

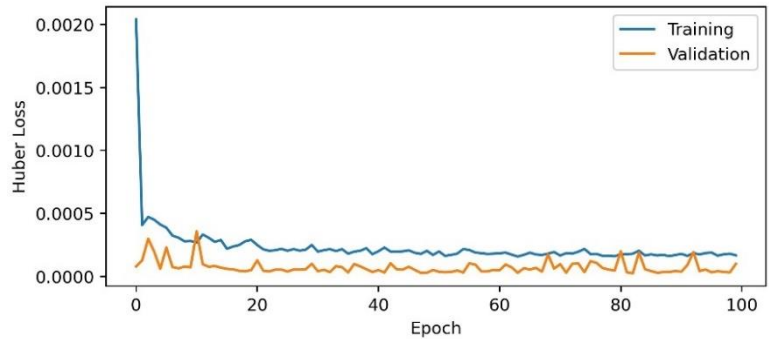

**Figure 14.** The loss of the LSTM model for the ZEEL stock

The training and validation loss for the LSTM model for Zee Entertainment Enterprises, the leading stock of this sector, are plotted in Figure 14. The actual and predicted prices of the Zee Entertainment Enterprises stock by the LSTM model from January 1, 2024, to April 23, 2024, are depicted in Figure 15.

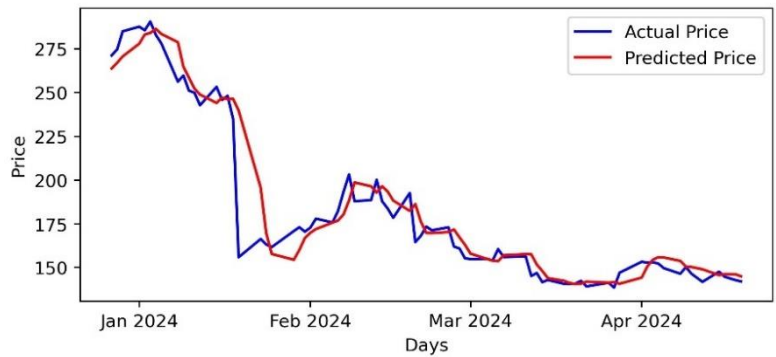

**Figure 15.** The actual vs predicted price of the ZEEL stock from January 1, 2024, to April 23, 2024
.
**Table 8.** The results of the Metal sector stocks (Period: January 1, 2005 – April 23, 2024)

| Stock | Buy Profit | Sell Profit | Gross Profit | Mean Price | Gross Profit / Mean Price |
|---|---|---|---|---|---|
| TATASTEEL | 21913 | 23455 | 45368 | 33 | 1374.76 |
| ADANIENT | 305161 | 307647 | 612808 | 357 | 1716.55 |
| HINDALCO | 82848 | 82756 | 165604 | 149 | 1111.44 |
| JSWSTEEL | 111850 | 116894 | 228744 | 135 | 1694.40 |
| VEDL | 50943 | 50928 | 101871 | 65 | 1567.25 |
| JINDALSTEL | 140260 | 140621 | 280881 | 229 | 1226.55 |
| APLAPOLLO | 128784 | 132225 | 261009 | 340 | 767.67 |
| JSL | 421889 | 39954 | 82143 | 103 | 797.50 |
| NMDC | 13530 | 12546 | 26076 | 71 | 367.27 |
| SAIL | 33863 | 33886 | 67749 | 54 | 1254.61 |
| **Sectoral average of gross profit/mean price** | | | | | **1187.80** |

*Metal sector:* Based on the NSE's report published on March 28, 2024, the top ten stocks of this sector based on their free-float market capitalization values and their respective weights (in percentage) in the computation of the overall sectoral index are as follows. Tata Steel (TATASTEEL): 22.23, Adani Enterprises (ADANIENT): 14.51, Hindalco Industries (HINDALCO): 14.17, JSW Steel (JSWSTEEL): 13.71, Vedanta (VEDL): 6.29, Jindal Steel & Power (JINDALSTEL): 5.40, APL Apollo Tubes (APLAPOLLO): 3.73, Jindal Stainless (JSL): 3.74, NMDC (NMDC): 3.99, and Steel Authority of India (SAIL): 3.36 (NSE Website). The results of this sector are presented in Table 8.

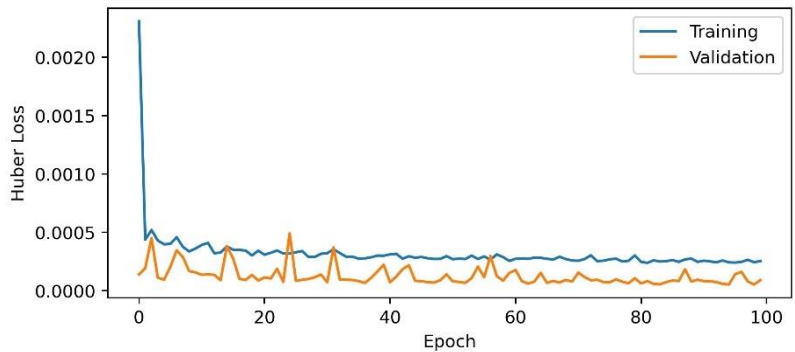

**Figure 16.** The loss of the LSTM model for the TATASTEEL stock

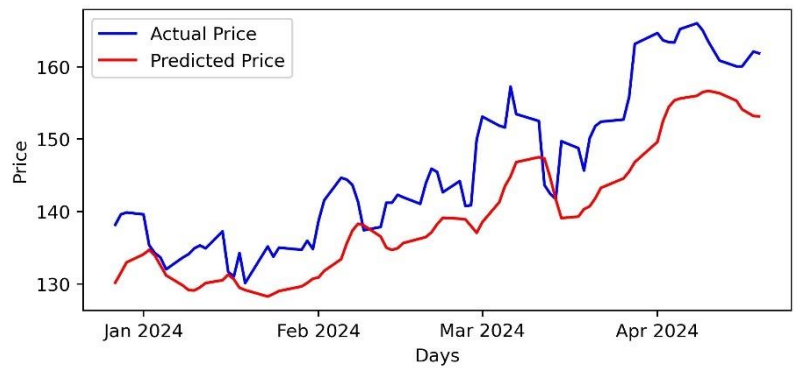

**Figure 17.** The actual vs predicted price of the TATASTEEL stock from January 1, 2024, to April 23, 2024

The training and validation loss for the LSTM model for Tata Steel, the leading stock of this sector, are plotted in Figure 16. The actual and predicted prices of the Tata Steel stock by the LSTM model from January 1, 2024, to April 23, 2024, are depicted in Figure 17.

*Oil & Gas sector:* Based on the NSE's report published on March 28, 2024, the top ten stocks of this sector based on their free-float market capitalization values and their respective weights (in percentage) in the computation of the overall sectoral index are as follows. Reliance Industries (RELIANCE): 33.05, Oil & Natural Gas Corporation (ONGC): 16.41, Indian Oil Corporation (IOC): 9.67, Bharat Petroleum Corporation (BPCL): 9.03, GAIL India (GAIL): 7.66, Hindustan Petroleum Corporation (HINDPETRO): 4.77, Adani Total Gas (ATGL): 4.00, Oil India (OIL): 3.37, Petronet LNG (PETRONET): 3.10, and Indraprastha Gas (IGL): 2.37 (NSE Website). The results of this sector are presented in Table 9.

The training and validation loss for the LSTM model for Reliance Industries, the leading stock of this sector, are plotted in Figure 18. The actual and predicted prices of the Reliance

Industries stock by the LSTM model from January 1, 2024, to April 23, 2024, are depicted in Figure 19.

**Table 9.** The results of the Oil & Gas sector stocks (Period: January 1, 2005 – April 23, 2024)

| Stock | Buy Profit | Sell Profit | Gross Profit | Mean Price | Gross Profit / Mean Price |
|---|---|---|---|---|---|
| RELIANCE | 5145450 | 503801 | 1018351 | 594 | 1714.40 |
| ONGC | 42069 | 42523 | 84592 | 78 | 1084.51 |
| IOC | 21248 | 19707 | 40955 | 31 | 1321.13 |
| BPCL | 103891 | 103585 | 207476 | 127 | 1633.67 |
| GAIL | 25831 | 25168 | 50999 | 43 | 1159.07 |
| HINDPETRO | 70015 | 70659 | 140674 | 96 | 1465.35 |
| ATGL | 140428 | 146760 | 287188 | 1043 | 275.35 |
| OIL | 21666 | 20384 | 42050 | 151.29 | 278.48 |
| PETRONET | 45691 | 41505 | 87196 | 97.25 | 898.93 |
| IGL | 86828 | 83860 | 170688 | 163 | 1047.17 |
| **Sectoral average of gross profit/mean price** | | | | | **1087.81** |

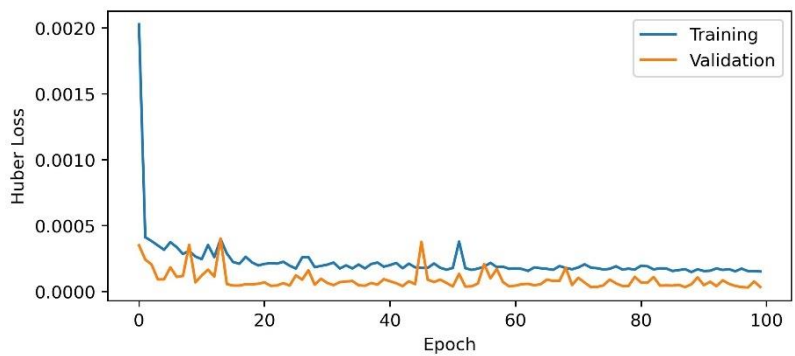

**Figure 18.** The loss of the LSTM model for the RELIANCE stock

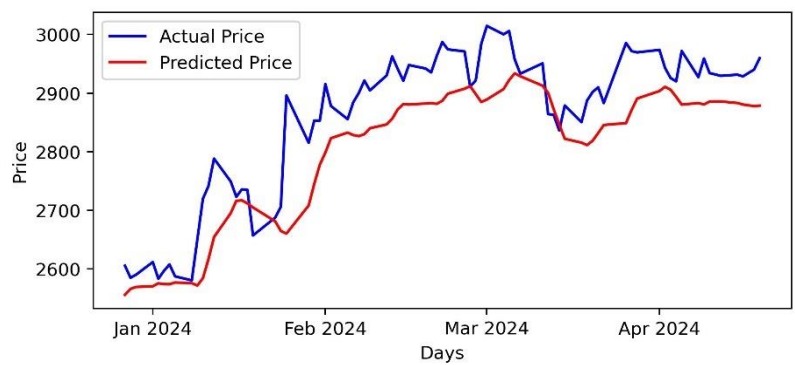

**Figure 19.** The actual vs predicted price of the RELIANCE stock from January 1, 2024, to April 23, 2024

*Realty sector:* Based on the NSE's report published on March 28, 2024, the top ten stocks of this sector based on their free-float market capitalization values and their respective weights (in percentage) in the computation of the overall sectoral index are as follows. DLF (DLF):

29.27, Macrotech Developers (LODHA): 15.46, Godrej Properties (GODREJPROP): 13.29, Phoenix Mills (PHOENIXLTD): 13.11, Oberoi Realty (OBEROIRLTY): 8.71, Prestige Estates Projects (PRESTIGE): 8.32, Brigade Enterprises (BRIGADE): 5.92, Sobha (SOBHA): 2.65, Mahindra Lifespace Developers (MAHLIFE): 2.21, and Sunteck Realty (SUNTECK): 1.07 (NSE Website). The results of this sector are presented in Table 10.

**Table 10.** The results of the Realty sector stocks (Period: January 1, 2005 – APRIL 23, 2024)

| Stock | Buy Profit | Sell Profit | Gross Profit | Mean Price | Gross Profit / Mean Price |
|---|---|---|---|---|---|
| DLF | 61496 | 61290 | 122786 | 249 | 493.12 |
| LODHA | 16496 | 15269 | 31765 | 635 | 50.02 |
| GODREJPROP | 186442 | 186248 | 372690 | 709 | 525.66 |
| PHOENIXLTD | 184569 | 188196 | 372765 | 527 | 707.33 |
| OBEROIRLTY | 60519 | 58360 | 118879 | 377 | 315.33 |
| PRESTIGE | 67335 | 60273 | 127608 | 283 | 450.91 |
| BRIGADE | 69151 | 69796 | 138947 | 172 | 807.84 |
| SOBHA | 94173 | 96158 | 190331 | 394 | 483.07 |
| MAHLIFE | 55237 | 57570 | 112807 | 138 | 817.44 |
| SUNTECK | 48794 | 49142 | 97936 | 264 | 370.97 |
| **Sectoral average of gross profit/mean price** | | | | | **502.17** |

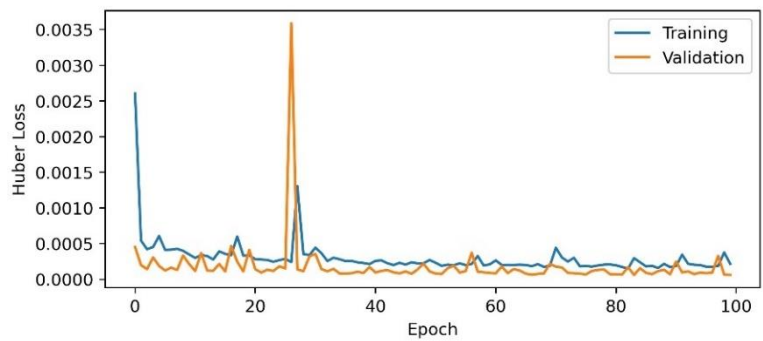

**Figure 20.** The loss of the LSTM model for the DLF stock

The training and validation loss for the LSTM model for DLF, the leading stock of this sector, are plotted in Figure 20. The actual and predicted prices of the DLF stock by the LSTM model for the period January 1, 2024, to April 23, 2024, are depicted in Figure 21.

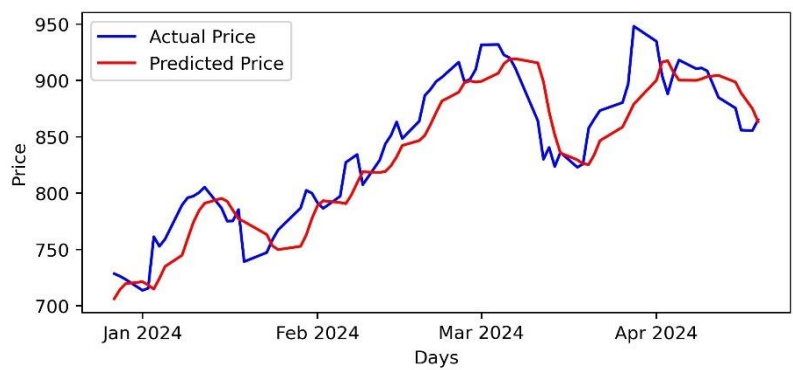

**Figure 21.** The actual vs predicted price of the DLF stock from January 1, 2024, to April 23, 2024

**Table 11.** The results of the Infrastructure sector stocks (Period: January 1, 2005 – April 23, 2024)

| Stock | Buy Profit | Sell Profit | Gross Profit | Mean Price | Gross Profit / Mean Price |
|---|---|---|---|---|---|
| RELIANCE | 514550 | 503801 | 1018351 | 594 | 1714.40 |
| LT | 388604 | 368859 | 757463 | 808 | 937.45 |
| BHATIARTL | 120277 | 116900 | 237177 | 334 | 710.11 |
| NTPC | 19020 | 17536 | 36556 | 100 | 365.56 |
| POWERGRID | 19221 | 18617 | 37838 | 72 | 525.53 |
| ULTRACEMCO | 1302162 | 1291108 | 2593270 | 2688 | 964.76 |
| ONGC | 42069 | 42523 | 84592 | 78 | 1084.51 |
| ADANIPORTS | 100305 | 102158 | 202463 | 337 | 600.78 |
| GRASIM | 280449 | 277249 | 557698 | 606 | 920.29 |
| TATAPOWER | 50317 | 47505 | 97822 | 67 | 1460.03 |
| **Sectoral average of gross profit/mean price** | | | | | **928.34** |

*Infrastructure sector:* Based on the NSE's report published on March 28, 2024, the top ten stocks of this sector based on their free-float market capitalization values and their respective weights (in percentage) in the computation of the overall sectoral index are as follows. Reliance Industries (RELIANCE): 19.97, Larsen & Toubro (LT): 151.51, Bharti Airtel (BHARTIARTL): 11.14, NTPC (NTPC): 5.56, Power Grid Corporation of India (POWERGRID): 4.40, Ultra Tech Cement (ULTRACEMCO): 3.92, Oil & Natural Gas Corporation (ONGC): 3.64, Adani Ports and Special Economic Zone (ADANIPORTS): 3.44, Grasim Industries (GRASIM): 2.94, and Tata Power Company (TATAPOWER): 2.33 (NSE Website). The results of this sector are presented in Table 11.

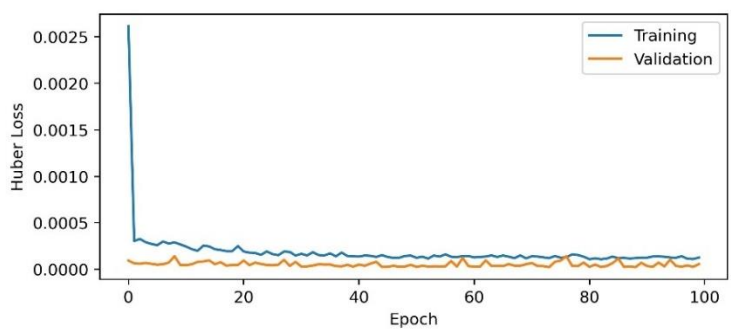

**Figure 22.** The loss of the LSTM model for the LT stock

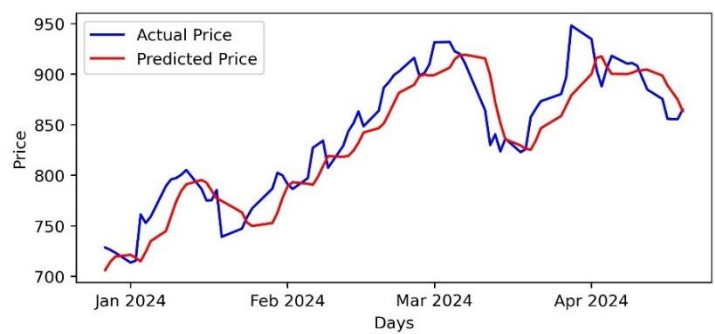

**Figure 23.** The actual vs predicted price of the LT stock from January 1, 2024, to April 23, 2024

The training and validation loss for the LSTM model for Larsen & Toubro, the stock with the second highest market capitalization in the realty sector, are plotted in Figure 22. The stock of Larsen & Toubro is chosen as the plots for Reliance Industries have already been shown in the *oil & gas* sector. The actual and predicted prices of the Larsen & Toubro stock by the LSTM model from January 1, 2024, to April 23, 2024, are depicted in Figure 23.

**Table 12.** The results of the Energy sector stocks (Period: January 1, 2005 – April 23, 2024)

| Stock | Buy Profit | Sell Profit | Gross Profit | Mean Price | Gross Profit / Mean Price |
|---|---|---|---|---|---|
| RELIANCE | 514550 | 503801 | 1018351 | 594 | 1714.40 |
| NTPC | 19020 | 17536 | 36556 | 100 | 365.56 |
| POWERGRID | 19221 | 18617 | 37838 | 72 | 525.53 |
| ONGC | 42069 | 42523 | 84592 | 78 | 1084.51 |
| COALINDIA | 16451 | 15990 | 32441 | 156 | 207.96 |
| ADANIGREEN | 111356 | 115630 | 226986 | 924 | 245.66 |
| TATAPOWER | 50317 | 47505 | 97822 | 67 | 1460.03 |
| IOC | 21248 | 19707 | 40955 | 31 | 1321.13 |
| BPCL | 103891 | 103585 | 207476 | 127 | 1633.67 |
| ADANIENSOL | 1694 | 1379 | 3073 | 966 | 3.18 |
| **Sectoral average of gross profit/mean price** | | | | | **856.16** |

*Energy sector:* Based on the NSE's report published on March 28, 2024, the top ten stocks of this sector based on their free-float market capitalization values and their respective weights (in percentage) in the computation of the overall sectoral index are as follows. Reliance Industries (RELIANCE): 33.17, NTPC (NTPC): 13.80, Power Grid Corporation (POWERGRID): 10.91, Oil & Natural Gas Corporation (ONGC): 9.04, Coal India (COALINDIA): 8.56, Adani Green Energy (ADANIGREEN): 5.78, Tata Power Company (TATAPOWER): 5.77, 9.94, Indian Oil Corporation (IOC): 5.33, Bharat Petroleum Corporation (BPCL): 4.97, and Adani Energy Solutions (ADANIENSOL): 2.67 (NSE Website). The results of this sector are presented in Table 12.

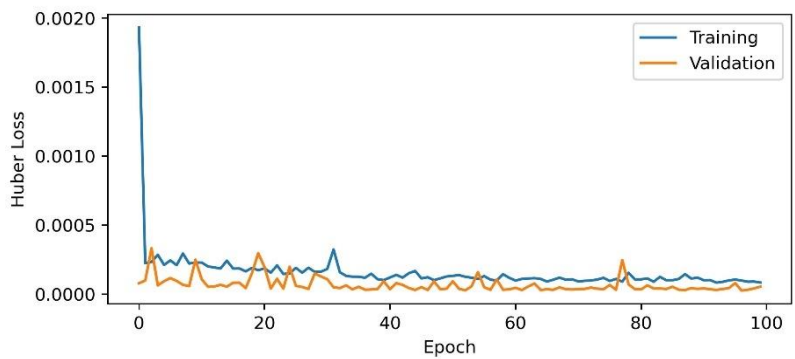

**Figure 24.** The loss of the LSTM model for the NTPC stock

The training and validation loss for the LSTM model for NTPC, the stock with the second-highest market capitalization in the *energy* sector, are plotted in Figure 24. The stock of NTPC is chosen as the plots for Reliance Industries have already been shown in the *oil & gas* sector. The actual and predicted prices of the NTPC stock by the LSTM model from January 1, 2024, to April 23, 2024, are depicted in Figure 25.

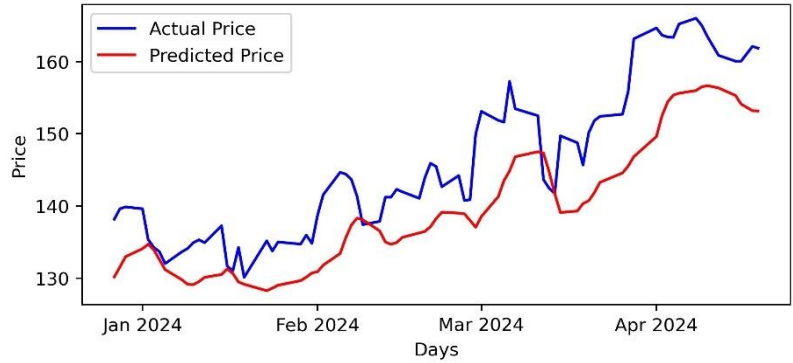

**Figure 25.** The actual vs predicted price of the NTPC stock from January 1, 2024, to April 23, 2024

*Private Banks sector:* Based on the NSE's report published on March 28, 2024, the top ten stocks of this sector based on their free-float market capitalization values and their respective weights (in percentage) in the computation of the overall sectoral index are as follows. HDFC Bank (HDFCBANK): 25.83, ICICI Bank (ICICIBANK): 25.81, IndusInd Bank (INDUSINDBK): 10.54, Axis Bank (AXISBANK): 10.40, Kotak Mahindra Bank (KOTAKBANK): 10.31, Federal Bank (FEDERALBNK): 5.63, IDFC First Bank (IDFCFIRSTB): 4.99, Bandhan Bank (BANDHANBNK): 2.47, RBL Bank (RBLBANK): 2.41, and City Union Bank (CUB): 1.62 (NSE Website). The results of this sector are presented in Table 13.

**Table 13.** The results of the Private Banks sector stocks (Period: January 1, 2005 – April 23, 2024)

| Stock | Buy Profit | Sell Profit | Gross Profit | Mean Price | Gross Profit / Mean Price |
|---|---|---|---|---|---|
| HDFCBANK | 364131 | 370223 | 734354 | 433 | 1695.97 |
| ICICIBANK | 140462 | 132266 | 272728 | 271 | 1006.38 |
| INDUSINDBK | 337654 | 330735 | 668389 | 571 | 1170.56 |
| AXISBANK | 208270 | 209023 | 417293 | 315 | 1324.74 |
| KOTAKBANK | 401103 | 399902 | 801005 | 651 | 1230.42 |
| FEDERALBNK | 29606 | 29812 | 59418 | 39 | 1523.54 |
| IDFCFIRSTB | 3127 | 3374 | 6501 | 51 | 127.47 |
| BANDHANBNK | 16824 | 19142 | 35966 | 339 | 106.09 |
| RBLBANK | 31535 | 33941 | 65476 | 312 | 209.86 |
| CUB | 39928 | 39077 | 79005 | 70 | 1145.00 |
| **Sectoral average of gross profit/mean price** | | | | | **954.00** |

The training and validation loss for the LSTM model for ICICI Bank, the stock with the second highest market capitalization in the *private banks* sector, are plotted in Figure 26. The stock of ICICI Bank is chosen as the plots for HDFC Bank (the stock with the highest market capitalization) have already been shown in the analysis of the *banking* sector. The actual and predicted prices of the ICICI Bank stock by the LSTM model from January 1, 2024, to April 23, 2024, are depicted in Figure 27.

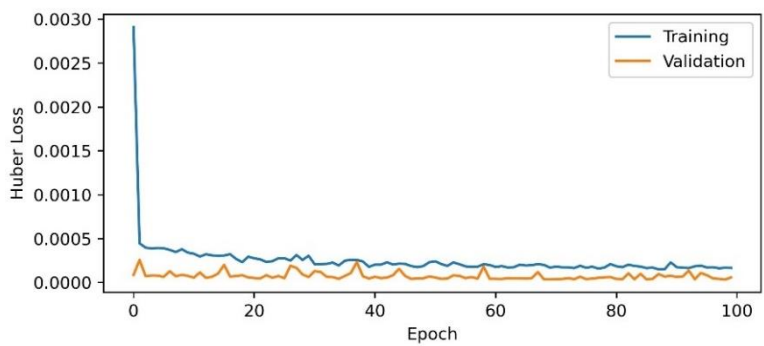

**Figure 26.** The loss of the LSTM model for the ICICIBANK stock

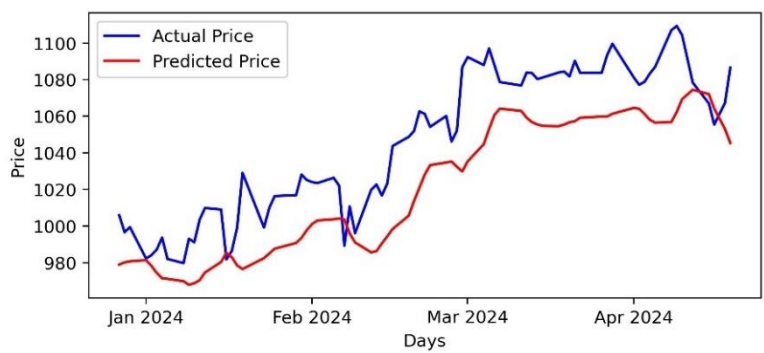

**Figure 27.** The actual vs predicted price of the ICICIBANK stock from January 1, 2024, to April 23, 2024
.

*PSU Banks sector:* Based on the NSE's report published on April 23, 2024, the top ten stocks of this sector based on their free-float market capitalization values and their respective weights (in percentage) in the computation of the overall sectoral index are as follows. State Bank of India (SBIN): 32.69, Bank of Baroda (BANKBARODA): 15.92, Canara Bank (CANBK): 12.63, Punjab National Bank (PNB): 11.98, Union Bank of India (UNIONBANK): 9.49, Indian Bank (INDIANB): 5.91, Bank of India (BANKINDIA): 5.46, Bank of Maharashtra (MAHABANK): 2.00, Indian Overseas Bank (IOB): 1.47, and Central Bank of India (CENTRALBK): 1.18 (NSE Website). The results of this sector are presented in Table 14.

**Table 14.** The results of the PSU Banks sector stocks (Period: January 1, 2005 – April 23, 2024)

| Stock | Buy Profit | Sell Profit | Gross Profit | Mean Price | Gross Profit / Mean Price |
|---|---|---|---|---|---|
| SBIN | 120565 | 118709 | 239274 | 175 | 1367.28 |
| BANKBARODA | 34257 | 32601 | 66858 | 95 | 703.77 |
| CANBK | 69171 | 63158 | 132329 | 231 | 572.85 |
| PNB | 30871 | 32024 | 62895 | 91 | 691.15 |
| UNIONBANK | 37597 | 38271 | 75868 | 108 | 702.48 |
| INDIANB | 37838 | 38220 | 76058 | 163 | 466.61 |
| BANKINDIA | 56999 | 60023 | 117022 | 145 | 807.04 |
| MAHABANK | 6951 | 6886 | 13837 | 28 | 494.18 |
| IOB | 19988 | 20728 | 40726 | 49 | 831.14 |
| CENTRALBK | 16110 | 16346 | 32456 | 64 | 507.13 |
| **Sectoral average of gross profit/mean price** | | | | | **714.36** |

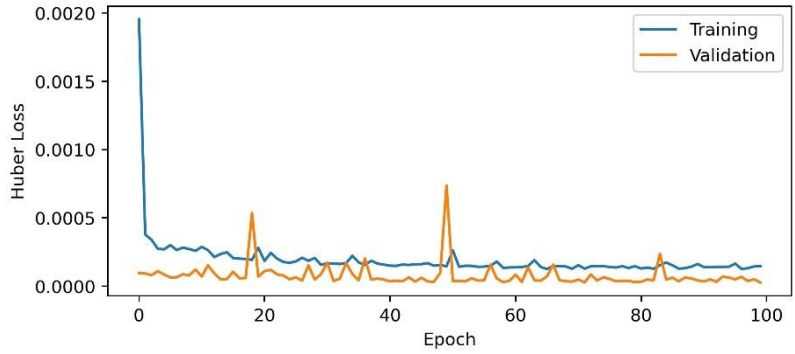

**Figure 28.** The loss of the LSTM model for the SBIN stock

The training and validation loss for the LSTM model for the State Bank of India (SBI), the stock with the highest market capitalization in the *PSU banks* sector, are plotted in Figure 28. The actual and predicted prices of the SBIN stock by the LSTM model for the period January 1, 2024, to April 23, 2024, are depicted in Figure 29.

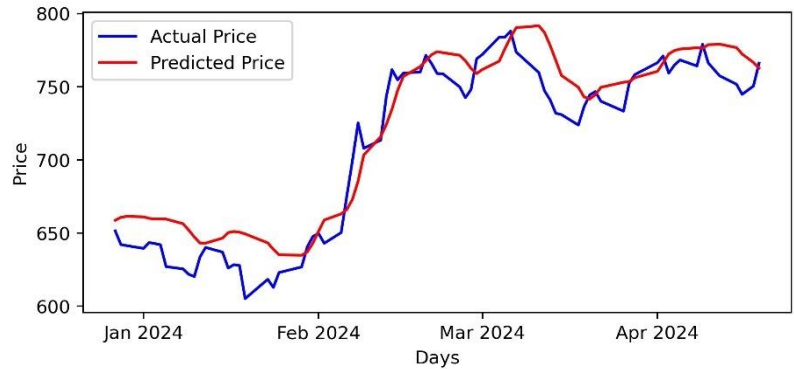

**Figure 29.** The actual vs predicted price of the SBIN stock from January 1, 2024, to April 23, 2024

**Table 15.** The results of the Consumption sector stocks (Period: January 1, 2005 – April 23, 2024)

| Stock | Buy Profit | Sell Profit | Gross Profit | Mean Price | Gross Profit / Mean Price |
|---|---|---|---|---|---|
| ITC | 82670 | 84128 | 166798 | 114 | 1463.15 |
| BHARTIARTL | 120277 | 116900 | 237177 | 334 | 710.11 |
| HINDUNILVR | 548689 | 561014 | 1109703 | 698 | 1589.83 |
| M&M | 288537 | 276382 | 564919 | 369 | 1530.95 |
| MARUTI | 1768834 | 1775983 | 3544817 | 3578 | 990.73 |
| TITAN | 498158 | 510217 | 1008375 | 505 | 1996.78 |
| ASIANPAINT | 563199 | 537224 | 1100423 | 865 | 1272.17 |
| ZOMATO | 3300 | 3135 | 6435 | 95 | 67.74 |
| BAJAJ-AUTO | 819358 | 833160 | 1652518 | 1647 | 1003.35 |
| NESTLEIND | 2806609 | 2951173 | 5757782 | 5518 | 1043.45 |
| **Sectoral average of gross profit/mean price** | | | | | **1166.83** |

*Consumption sector:* Based on the NSE's report published on April 23, 2024, the top ten stocks of this sector based on their free-float market capitalization values and their respective weights (in percentage) in the computation of the overall sectoral index are as follows. ITC (ITC): 9.85, Bharti Airtel (BHARTIARTL): 9.80, Hindustan Unilever (HINDUNILVR): 7.41, Mahindra & Mahindra (M&M): 6.74, Marui Suzuki (MARUTI): 6.10, Titan Company (TITAN): 5.81, Asian Paints (ASIANPAINT): 4.70, Zomato (ZOMATO): 3.88, Bajaj Auto (BAJAJ-AUTO): 3.80, and Nestle India (NESTLEIND): 3.43 (NSE Website). The results of this sector are presented in Table 15.

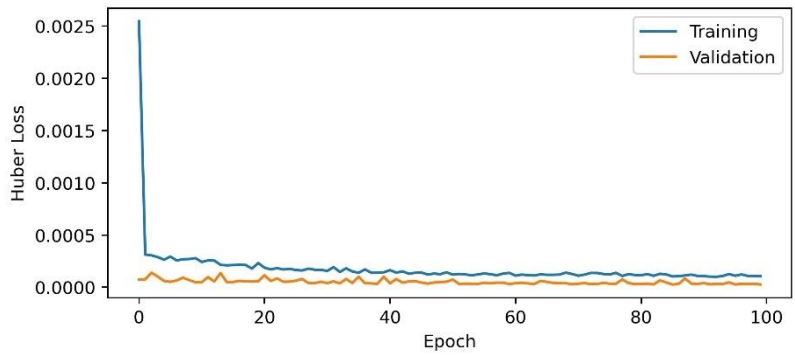

**Figure 30.** The loss of the LSTM model for the BHARTIARTL stock

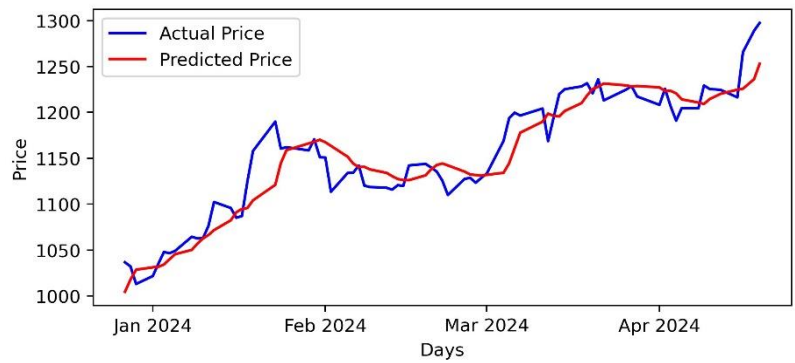

**Figure 31.** The actual vs predicted price of the BHARTIARTL stock from January 1, 2024, to April 23, 2024

The training and validation loss for the LSTM model for Bharti Airtel (BHARTIARTL), the stock with the second-highest market capitalization in the *consumption* sector, are plotted in Figure 30. The actual and predicted prices of the BHARTIARTL stock by the LSTM model for the period January 1, 2024, to April 23, 2024, are depicted in Figure 31. It may be noted that the plots for the stock with the highest free-float market capitalization in this sector, ITC, have been shown under the FMCG sector analysis.

*ESG sector:* Based on the NSE's report published on April 23, 2024, the top ten stocks of this sector based on their free-float market capitalization values and their respective weights (in percentage) in the computation of the overall sectoral index are as follows. Infosys (INFY): 5.99, HDFC Bank (HDFCBANK): 5.47, ICICI Bank (ICICIBANK): 4.05, Axis Bank (AXISBANK): 3.35, Kotak Mahindra Bank (KOTAKBANK): 3.30, Bharti Airtel (BHARTIARTL): 2.90, HCL Technologies (HCLTECH): 2.90, Tata Motors

(TATAMOTORS): 2.78, Tata Consultancy Services (TCS): 2.63, and Reliance Industries (RELIANCE): 2.21 (NSE Website). The results of this sector are presented in Table 16.

**Table 16.** The results of the ESG sector stocks (Period: January 1, 2005 – April 23, 2024)

| Stock | Buy Profit | Sell Profit | Gross Profit | Mean Price | Gross Profit / Mean Price |
|---|---|---|---|---|---|
| INFY | 295161 | 308620 | 603781 | 382 | 1580.58 |
| HDFCBANK | 364131 | 370223 | 734354 | 433 | 1695.97 |
| ICICIBANK | 140462 | 132266 | 272728 | 271 | 1006.38 |
| AXISBANK | 208270 | 209023 | 417293 | 315 | 1324.74 |
| KOTAKBANK | 401103 | 399902 | 801005 | 651 | 1230.42 |
| BHARTIARTL | 120277 | 116900 | 237177 | 334 | 710.11 |
| HCLTECH | 205669 | 194262 | 399931 | 318 | 1257.64 |
| TATAMOTORS | 163784 | 162912 | 326696 | 180 | 1814.98 |
| TCS | 610753 | 611463 | 1222216 | 1093 | 1118.22 |
| RELIANCE | 5145450 | 503801 | 1018351 | 594 | 1714.40 |
| **Sectoral average of gross profit/mean price** | | | | | **1345.34** |

The training and validation loss for the LSTM model for Axis Bank (AXISBANK), the stock with the fourth-highest market capitalization in the *consumption* sector, are plotted in Figure 32. The actual and predicted prices of the AXISBANK stock by the LSTM model for the period January 1, 2024, to April 23, 2024, are depicted in Figure 33. The plots for INFY, HDFCBANK, and ICICIBANK, the top three stocks in this sector have already been exhibited under the IT, Banking, and Private Banks sectors, respectively.

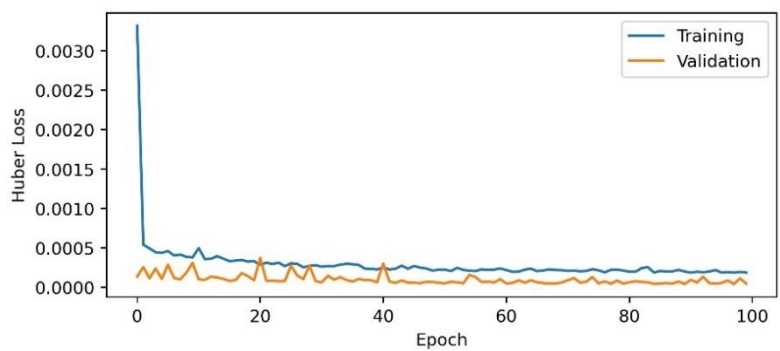

**Figure 32.** The loss of the LSTM model for the AXISBANK stock

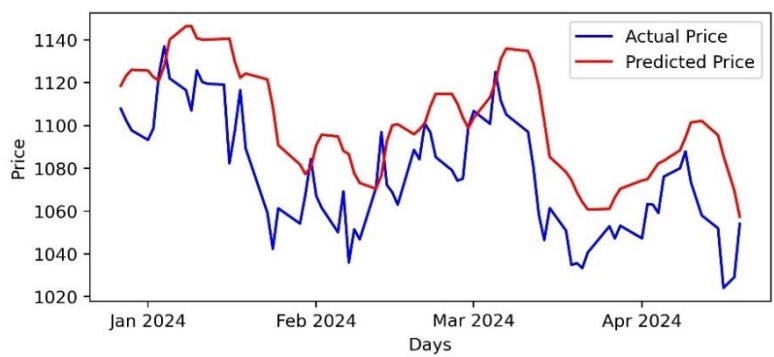

**Figure 33.** The actual vs predicted price of the AXISBANK stock from January 1, 2024, to April 23, 2024

*Financial Services sector:* Based on the NSE's report published on April 23, 2024, the top ten stocks of this sector based on their free-float market capitalization values and their respective weights (in percentage) in the computation of the overall sectoral index are as follows. HDFC Bank (HDFCBANK): 30.85, ICICI Bank (ICICIBANK): 21.75, Axis Bank (AXISBANK): 8.42, State Bank of India (SBIN): 8.18, Kotak Mahindra Bank (KOTAKBANK): 7.44, Bajaj Finance (BAJFINANCE): 5.72, Bajaj Finserv (BAJAJFINSV): 2.53, SBI Life Insurance Company (SBILIFE): 1.92, HDFC Life Insurance Company (HDFCLIFE): 1.89, Shriram Finance (SHRIRAMFIN): 1.86 (NSE Website). The results of this sector are presented in Table 17.

**Table 17.** The results of the financial services sector stocks (Period: January 1, 2005 – April 23, 2024)

| Stock | Buy Profit | Sell Profit | Gross Profit | Mean Price | Gross Profit / Mean Price |
|---|---|---|---|---|---|
| HDFCBANK | 364131 | 370223 | 734354 | 433 | 1695.97 |
| ICICIBANK | 140462 | 132266 | 272728 | 271 | 1006.38 |
| AXISBANK | 208270 | 209023 | 417293 | 315 | 1324.74 |
| SBIN | 120565 | 118709 | 239274 | 175 | 1367.28 |
| KOTAKBANK | 401103 | 399902 | 801005 | 651 | 1230.42 |
| BAJFINANCE | 1280890 | 1215891 | 2496781 | 1546 | 1614.99 |
| BAJAJFINSV | 446292 | 437363 | 883655 | 775 | 1140.20 |
| SBILIFE | 46837 | 47402 | 94239 | 994 | 94.81 |
| HDFCLIFE | 17269 | 16080 | 33349 | 564 | 59.13 |
| SHRIRAMFIN | 298916 | 293856 | 592772 | 626 | 946.92 |
| **Sectoral average of gross profit/mean price** | | | | | **1048.08** |

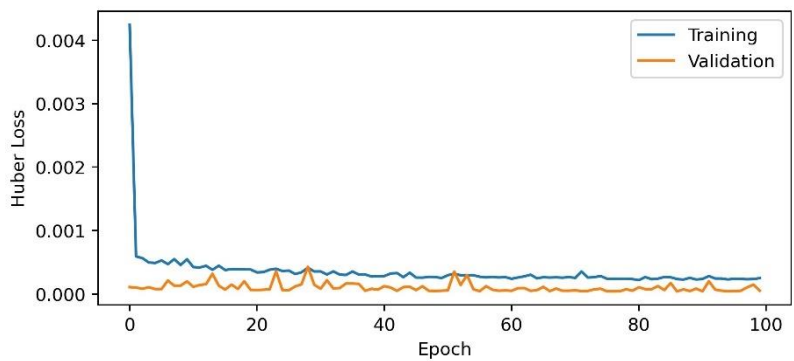

**Figure 34.** The loss of the LSTM model for the KOTAKBANK stock

The training and validation loss for the LSTM model for Kotak Mahindra Bank (KOTAKBANK), the stock with the fifth-highest market capitalization in the *financial services* sector, are plotted in Figure 34. The actual and predicted prices of the KOTAKBANK stock by the LSTM model for the period January 1, 2024, to April 23, 2024, are depicted in Figure 35. The plots for HDFCBNAK, ICICIBANK, AXISBANK, and SBIN, the top four stocks in this sector have already been exhibited under the Banking, Private Banks, ESG, and PSU Banks sectors, respectively.

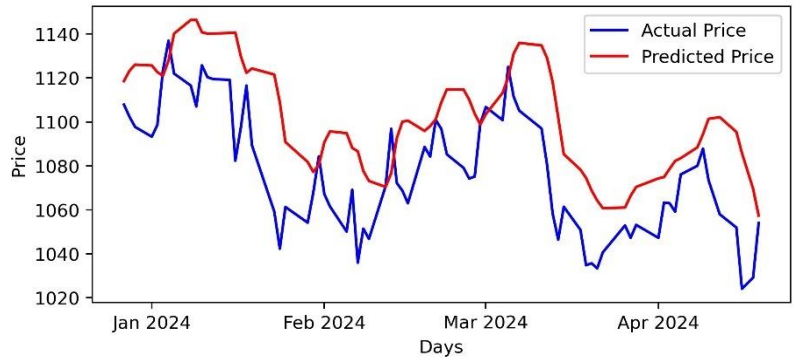

**Figure 35.** The actual vs predicted price of the KOTAKBANK stock from January 1, 2024, to April 23, 2024

**Table 18.** The results of the services sector stocks (Period: January 1, 2005 – April 23, 2024)

| Stock | Buy Profit | Sell Profit | Gross Profit | Mean Price | Gross Profit / Mean Price |
|---|---|---|---|---|---|
| HDFCBANK | 364131 | 370223 | 734354 | 433 | 1695.97 |
| ICICIBANK | 140462 | 132266 | 272728 | 271 | 1006.38 |
| INFY | 295161 | 308620 | 603781 | 382 | 1580.58 |
| TCS | 610753 | 611463 | 1222216 | 1093 | 1118.22 |
| BHARTIARTL | 120277 | 116900 | 237177 | 334 | 710.11 |
| AXISBANK | 208270 | 209023 | 417293 | 315 | 1324.74 |
| SBIN | 120565 | 118709 | 239274 | 175 | 1367.28 |
| KOTAKBANK | 401103 | 399902 | 801005 | 651 | 1230.42 |
| BAJFINANCE | 1280890 | 1215891 | 2496781 | 1546 | 1614.99 |
| HCLTECH | 205669 | 194262 | 399931 | 318 | 1257.64 |
| **Sectoral average of gross profit/mean price** | | | | | **1290.63** |

*Services sector:* Based on the NSE's report published on April 23, 2024, the top ten stocks of this sector based on their free-float market capitalization values and their respective weights (in percentage) in the computation of the overall sectoral index are as follows. HDFC Bank (HDFCBANK): 18.38, ICICI Bank (ICICIBANK): 12.96, Infosys (INFY): 9.03, Tata Consultancy Services (TCS): 6.63, Bharti Airtel (BHARTIARTL): 5.39, Axis Bank (AXISBANK): 5.02, State Bank of India (SBIN): 4.87, Kotak Mahindra Bank (KOTAKBANK): 4.43, Bajaj Finance (BAJFINANCE): 3.41, and HCL Technologies (HCLTECH): 2.76 (NSE Website). The results of this sector are presented in Table 18.

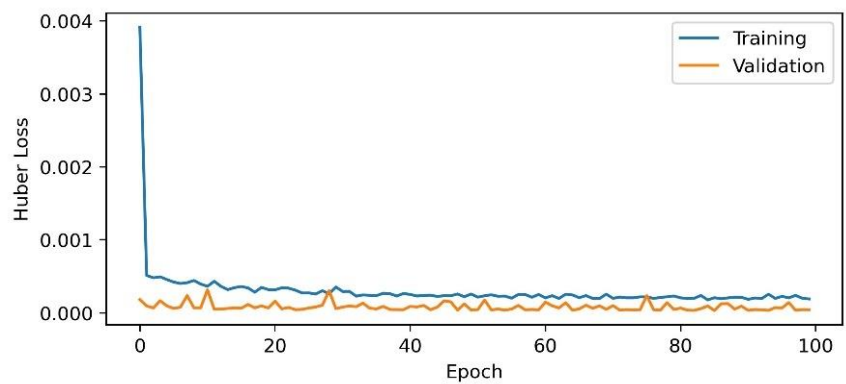

**Figure 36.** The loss of the LSTM model for the TCS stock

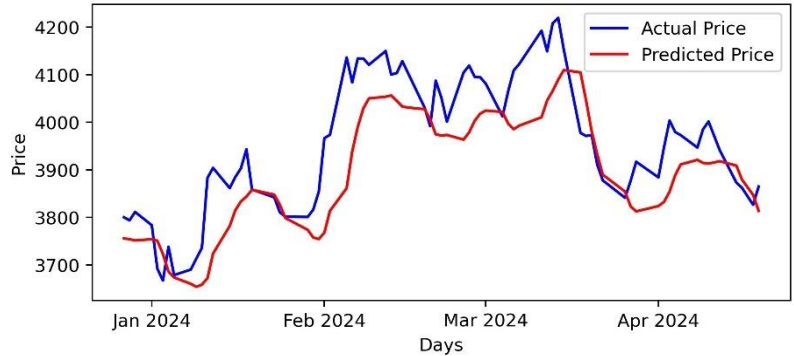

**Figure 37.** The actual vs predicted price of the TCS stock from January 1, 2024, to April 23, 2024

The training and validation loss for the LSTM model for Tata Consultancy Services (TCS), the stock with the fourth-largest market capitalization in the *services* sector, are plotted in Figure 36. The actual and predicted prices of the TCS stock by the LSTM model for the period January 1, 2024, to April 23, 2024, are depicted in Figure 37. The plots for HDFCBNAK, ICICIBANK, and INFY, the top three stocks in this sector have already been exhibited under the Banking, Private Banks, and IT sectors, respectively.

**Table 19.** The summary results of the sectors

| Sector | Profit/Mean Price |
|---|---|
| ESG | 1345.34 |
| Services | 1290.63 |
| Pharma | 1274.43 |
| FMCG | 1221.42 |
| Auto | 1189.24 |
| Metal | 1187.80 |
| Consumption | 1166.83 |
| Oil & Gas | 1087.81 |
| Banking | 1084.13 |
| Financial Services | 1048.08 |
| Information Technology (IT) | 975.20 |
| Private Banks | 954.00 |
| Infrastructure | 928.34 |
| Consumer Durables | 873.38 |
| Energy | 856.16 |
| PSU Banks | 714.36 |
| Media | 704.42 |
| Realty | 502.17 |

Table 19 presents the summary of the results for the eighteen sectors. Since the *mean ratio of the total profit earned to the mean value of the close prices of stocks* is a good indicator of the profitability of a sector, the sectors are compared on this metric. In Table 19, the column *Profitability Index* of a sector refers to the mean ratio of the total profit earned to the mean close values of the stocks for a given sector. It is observed that ESG, Services, Pharma, and FMCG are the fur sectors that have yielded the highest profitability index. On the other hand, Realty, Media, PSU Banks, and Energy, are the four least profitable sectors.

**Table 20.** The average performance level of the LSTM model for the sectors (Period: January 1, 2005 – April 23, 2024)

| Sector | Huber Loss | MAE | Acc. Score |
|---|---|---|---|
| Auto | **0.000049** | 54.817930 | 0.978721 |
| Banking | 0.000077 | **14.563779** | 0.980422 |
| Consumer Durables | 0.000198 | 147.481989 | 0.964944 |
| FMCG | 0.000069 | 50.468272 | 0.976239 |
| Pharma | 0.000056 | 136.806372 | 0.982991 |
| IT | 0.000099 | 170.288376 | 0.976950 |
| Media | 0.000281 | 198.083550 | 0.965951 |
| Metal | 0.000084 | 15.360568 | 0.974476 |
| Oil & Gas | 0.000088 | 25.526608 | 0.974761 |
| Realty | 0.000131 | 99.397466 | 0.979414 |
| Infrastructure | 0.000053 | 58.285192 | 0.977980 |
| Energy | 0.000253 | 98.285480 | 0.958315 |
| Private Banks | 0.000130 | 37.069798 | 0.978968 |
| PSU Banks | 0.000097 | 16.970538 | **0.984592** |
| Consumption | 0.000073 | 75.674647 | 0.972297 |
| ESG | 0.000057 | 21.541403 | 0.980075 |
| Financial Services | 0.000129 | 108.744428 | 0.978810 |
| Services | 0.000068 | 27.513052 | 0.977330 |

*Performance Evaluation of the LSTM Model:* In the final analysis, the LSTM model's performance is thoroughly assessed using three key metrics. Firstly, the Huber Loss (HL) is employed to measure the loss exhibited by the model when tested with out-of-sample data, essentially the test data. This loss metric provides valuable insights into how well the model generalizes to unseen data. Secondly, the Mean Absolute Error (MAE) is calculated, which represents the average of the absolute differences between the actual stock prices and the prices predicted by the LSTM model on the test data. MAE is a crucial measure of the model's accuracy, reflecting how close its predictions are to the actual prices. Lastly, the Accuracy Score (AS) is determined, indicating the percentage of correct predictions made by the LSTM model regarding the direction of movement (i.e., whether the stock price will increase or decrease) for the next day. This accuracy score is particularly important for investors as it assists them in making informed buy or sell decisions.

While the performance of the LSTM model is studied for all 140 stocks from the 14 sectors, for space constraints, we presented the sector-wise average performance of the model in Table 20. This table provides a comprehensive overview of how well the model performs in various sectors, allowing for a nuanced evaluation of its effectiveness and reliability across different market segments.

*Observations:* For a model to be accurate, it must have low values for Huber loss and MAE, and a high value for the accuracy score. It is observed that the LSTM model has yielded the minimum value of Huber loss for the *auto* sector, the minimum value of MAE for the *banking* sector, and the maximum value of accuracy score for the *PSU banks* sector. Hence, the model is found to be the most accurate for these three sectors. On the other hand, Huber loss and MAE are the maximum for the *media* sector, and accuracy score is the minimum for the *energy* sector. Therefore, the performance of the model has been the worst for *media* and *energy* sectors on the three metrics. Overall, the performance of the model has been highly accurate, since the lowest value of the accuracy score is found to be 0.958315. In other words, in the worst case, the model correctly predicted the direction of movement of the price (upward or downward) of the stock the next day in 95.83% of cases. Thus, the model can be reliably used in stock trading decisions.

# 6. Conclusion

This paper introduces an LSTM model designed to predict future stock prices, optimized with carefully crafted layers, and employing dropout regularization for regularization. Historical stock price data from 180 stocks across 18 sectors listed on the NSE, India, is collected from January 1, 2005, to April 23, 2024. The model forecasts stock prices for one day ahead, and based on these predictions, buy/sell decisions are made. The total profit from these transactions is normalized by the mean price of the stock over the entire period to gauge its profitability. The profitability of each stock within a sector is aggregated to determine the overall profitability index of the sector. Notably, the ESG sector emerges as the most profitable, while the *realty* sector exhibits the lowest profitability index. The LSTM model's accuracy is assessed using three metrics: Huber loss, mean absolute error (MAE), and accuracy score. The results indicate high accuracy across all analyzed stocks. The model is optimized with suitably designed layers and regularized using the *dropout regularization* method. The historical stock prices for 180 stocks from 18 sectors listed in NSE, India are extracted from the web from January 1, 2005, to April 23, 2024. The model is used for predicting future stock prices with a forecast horizon of 1 day, and based on the predicted output of the model buy/sell decisions of stocks are taken. The total profit earned from the buy/sell transactions for a stock is normalized by its mean price over the entire period to arrive at the profitability measure of the stock. The profitability figures of all stocks in a given sector are summed up to derive the overall *profitability index* of the sector. As a future scope of work, stocks from the world's leading stock exchanges will be included in the study to analyze the performance of the model in the portfolio design of international stocks.